\documentclass[12pt,a4paper]{article}

\usepackage{epsfig}
\usepackage{amssymb}

\newcommand{\be}{\begin{equation}}
\newcommand{\ee}{\end{equation}}
\newcommand{\bea}{\begin{eqnarray}}
\newcommand{\eea}{\end{eqnarray}}

\newcommand{\Pp}{{\mathcal P}}

\newcommand{\pd}{\partial}
\newcommand{\Tr}{\mathop{\rm Tr}\nolimits}

\begin{document}
\title{
\begin{flushright}
{\small SMI-5-00 }
\end{flushright}
\vspace{1cm}A Note on UV/IR for Noncommutative Complex Scalar Field}
\author{
I. Ya. Aref'eva${}^{\S}$, D. M. Belov${}^{\dag}$ and A. S.
Koshelev${}^{\dag}$ \\
\\${}^{\S}$ {\it  Steklov Mathematical Institute,}\\ {\it Gubkin st.8,
Moscow, Russia, 117966}\\ arefeva@genesis.mi.ras.ru\\
\\${}^{\dag}$
{\it Physical Department, Moscow State University, }\\ {\it
Moscow, Russia, 119899} \\ belov@orc.ru, kas@depni.npi.msu.su}

\date {~}
\maketitle
\begin{abstract}
Noncommutative quantum field theory of a complex scalar field is
considered. There is a two-coupling  noncommutative analogue of
$U(1)$-invariant quartic interaction $(\phi^*\phi)^2$, namely
$A\phi^*\star\phi\star\phi^*\star\phi+
B\phi^*\star\phi^*\star\phi\star\phi$. For arbitrary values of $A$
and $B$ the model is nonrenormalizable. However, it is one-loop
renormalizable in two special cases: $B=0$ and $A=B$.
Furthermore, in the case $B=0$ the model does not suffer from IR
divergencies at least at one-loop insertions level.
\end{abstract}

\newpage
\section{Introduction}

Recently, there is a renovation of the interest in noncommutative
quantum field theories (or field theories on noncommutative
space-time \cite{book,Mad}). As emphasized in \cite{SW}, the
important question is whether or not the noncommutative quantum
field theory  is well-defined. Note that one of earlier
motivations to consider noncommutative field  theories  is a hope
that it would be possible to avoid  quantum field theory
divergencies \cite{WZ,AVqp,Mad,filk,Kempf}. Now a commonly
accepted belief is that a theory on a noncommutative space is
renormalizable iff the corresponding commutative theory is
renormalizable. Results on one-loop renormalizability of
noncommutative gauge theory \cite{ren} and   two-loop
renormalizability of noncommutative scalar $\phi _4^4$ theory
\cite{ABK} as well as general considerations \cite{SB,Ch} support
this belief. In this paper we show that for more complicated
models this is not true.

 Note that renormalizability does not guarantee that the
theory is well-defined. There is a mixing of the UV and the IR
divergencies \cite{MMS}. In particular, multi one-loop
insertions in $\varphi^3$ theory \cite{MMS} and multi tadpole
insertions in $\varphi^4$ theory \cite{ABK} produce  infrared
divergencies. UV/IR mixing depends on the model.
The $U(1)$ noncommutative gauge theory does not
exhibit a mixing
of the UV and the IR dynamics\cite{Hay}. For further discussions see
\cite{AV-M}-\cite{Grosse}.

The IR behaviour of noncommutative theories is closely related
with an existence of a commutative limit of a noncommutative
quantum
theory under consideration. In particular, the IR behaviour of
noncommutative $\varphi^4_4$ theory
makes an existence of the commutative limit  impossible.

In this paper we consider noncommutative quantum field theories
 of complex scalar field
\cite{AV} whose commutative analogue  $(\phi^*\phi)^2$ is
renormalizable in four-dimensional case. There is a two-coupling
noncommutative analogue of $U(1)$-invariant quartic interaction
$(\phi^*\phi)^2$, namely $A\phi^*\star\phi\star\phi^*\star\phi+
B\phi^*\star\phi^*\star\phi\star\phi$. For arbitrary values of $A$
and $B$ the model is nonrenormalizable. However it is one-loop
renormalizable in two special cases: $B=0$ and $A=B$. Moreover,
in the case $B=0$ the model does not suffer from IR
divergencies at least at one-loop insertions level.

\section{The model}
Consider complex scalar field. There are only two noncommutative
structures that generalize a commutative quartic interaction
$(\phi^*\phi)^2$:
\begin{description}
  \item[(a)] $~~~~~\Tr \phi^*\star\phi\star\phi^*\star\phi$,
  \item[(b)] $~~~~~\Tr \phi^*\star\phi^*\star\phi\star\phi$,
\end{description}
where $\star$ is the Moyal product
$(f\star g)(x)=e^{i\xi\theta^{\mu\nu}\pd_{\mu}\otimes\pd_{\nu}}
f(x)\otimes g(x)$, $\xi$ is a deformation parameter, $\theta^{\mu\nu}$
is a nondegenerate skew-symmetric real constant matrix.
In the commutative case the quartic interaction
$(\phi^*\phi)^2$  is invariant under  local $U(1)$-transformations.
In the noncommutative
theory we can consider a "deformed" $U(1)$-symmetry ($U\star U^*=1$).
 One sees that only the structure (a)
is invariant under these transformations.
Using (a) and (b) we can construct an interaction
$$
V[\phi^*,\phi]=A\Tr \phi^*\star\phi\star\phi^*\star\phi
+ B\Tr \phi^*\star\phi^*\star\phi\star\phi=
$$
\be
(A-B)\Tr\phi^*\star\phi\star\phi^*\star\phi+
\frac{B}{2}\Tr([\phi^*,\phi]_{AM}\star[\phi^*,\phi]_{AM}),
\ee
where $[,]_{AM}$ is the Moyal antibracket $[f,g]_{AM}=f\star g+g\star f$.
The action of the theory is
\be
S=\int d^dx\left[\pd_{\mu}\phi^*\pd_{\mu}\phi+m^2\phi^*\phi\right]+
V[\phi^*,\phi].
\label{action}
\ee
Let us rewrite the interaction term in the Fourier components
and symmetrize it, i.e.
$$
V[\phi^*,\phi]=\frac{1}{(2\pi)^4}\int
dp_1\dots dp_4\delta(\sum p_i)\times
$$
\be
\times[A\cos(p_1\wedge
p_2+p_3\wedge p_4)+B\cos(p_1\wedge p_3)\cos(p_2\wedge p_4)]
\phi^*(p_1)\phi(p_2)\phi^*(p_3)\phi(p_4).
\ee

\section{One Loop}
\begin{figure}[h]
\begin{center}
\epsfig{file=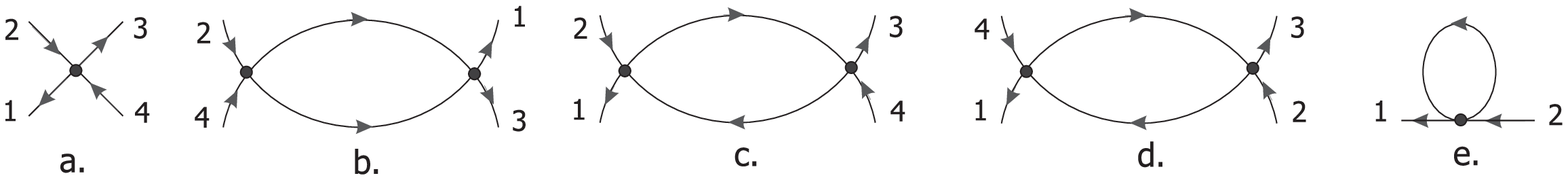,
   width=400pt,
  angle=0,
 }
\caption{The vertex and one-loop graphs} \label{F1}
\end{center}
\end{figure}

In this section we analyze counterterms to one loop Feynman graphs
in the  theory (\ref{action}) and find conditions when this theory
is renormalizable. All one-loop graphs are presented on
Fig.\ref{F1}:b,c,d. "In" arrows are the fields "$\phi$" and "out"
arrows are the fields "$\phi^*$".

The following analytic expression corresponds to the graph on
Fig.\ref{F1}:b
\be
\Gamma_{\ref{F1}b}=\frac{N_b}{(2\pi)^d}\int d^dk
\frac{\Pp_{\ref{F1}b}(p,k)}{(k^2+m^2)((k+P)^2+m^2)},
\label{1b}
\ee
where $N_b$ is a number of graphs ($N_b$=8), $P=p_2+p_4=-p_1-p_3$ and
$\Pp_{\ref{F1}b}(p,k)$ is the trigonometric polynomial
$$
\Pp_{\ref{F1}b}(p,k)=[A\cos(k\wedge p_2+(-k-p_2)\wedge
p_4)+B\cos(p_2\wedge p_4)\cos(k\wedge P)]
$$
\be
\times [A\cos(p_1\wedge(-k)+p_3\wedge (k-p_1))+B\cos(p_1\wedge p_3)\cos(k\wedge P)]
\ee
The terms containing $\exp[(\dots)\wedge k]$ give a finite contribution
to (\ref{1b}).
Divergencies come from the terms $\Delta\Pp_{\ref{F1}b}$ of the polynomial $\Pp_{\ref{F1}b}$
\be
\Delta\Pp_{\ref{F1}b}=\frac{B^2}{2}\cos(p_1\wedge p_3)
\cos(p_2\wedge p_4).
\ee

The graphs Fig.\ref{F1}:c and \ref{F1}:d mutually differ by
permutation of momenta
$1\leftrightarrow 3$ only and the analytic expressions
for these graphs coincide. For the graph Fig.\ref{F1}:c we have
\be
\Gamma_{\ref{F1}c}=\frac{N_c}{(2\pi)^d}\int d^dk
\frac{\Pp_{\ref{F1}c}(p,k)}{(k^2+m^2)((k+P)^2+m^2)},
\ee
where $N_c$ is a number of graphs ($N_c=16$), $P=p_1+p_2=-p_3-p_4$ and
$\Pp_{\ref{F1}c}(p,k)$ is the trigonometric polynomial
$$
\Pp_{\ref{F1}c}(p,k)=[A\cos(p_1\wedge p_2+(-k-P)\wedge
k)+B\cos(p_1\wedge (k+P))\cos(p_2\wedge k)]
$$
\be
\times [A\cos(p_3\wedge p_4+(-k)\wedge (k+P))+B\cos(p_3\wedge
k)\cos(4\wedge (k+P))].
\ee
The polynomial $\Delta\Pp_{\ref{F1}c}$ that gives contribution to a divergent
part of this graph is equal (after symmetrization
$p_2\leftrightarrow p_4$) to
\be \Delta \Pp_{\ref{F1}c}=\cos(p_1\wedge
p_2+p_3\wedge p_4) \left[\frac{A^2}{2}+\frac{B^2}{8}\right]+
\frac{AB}{2}\cos(p_1\wedge p_3) \cos(p_2\wedge
p_4).
\ee
We obtain the same answer for the graph on Fig.\ref{F1}:d, i.e.
$$
N_d=N_c,\qquad \Delta\Pp_{\ref{F1}c}=\Delta\Pp_{\ref{F1}d}.
$$

It is easy to see that the following condition is equal to one-loop
renormalizability of the theory (\ref{action})
\be
N_b\Delta\Pp_{\ref{F1}b}+2N_c\Delta\Pp_{\ref{F1}c}=C
[A\cos(p_1\wedge p_2+p_3\wedge p_4)+B\cos(p_1\wedge
p_3)\cos(p_2\wedge p_4)], \label{cond}
\ee

where $C$ is a constant.
The condition (\ref{cond}) yields two algebraic equations:
\bea
N_c\left[A^2+\frac{B^2}{4}\right]=AC\\
N_b\frac{B^2}{2}+N_cAB=BC
\eea
This system is self consistent if
$$
B(BN_c-2AN_b)=0.
$$
The last equation has two solutions: $B=0$ and $A=B$.
Therefore, one-loop renormalizability takes place only in two cases
\bea
B=0 & \mbox{and} & V[\phi^*,\phi]=A\Tr (\phi^*\star\phi)^2,\\
A=B & \mbox{and} & V[\phi^*,\phi]=\frac{B}{2}\Tr([\phi^*,\phi]_{AM})^2.
\eea

Theories with a real scalar field have problems with infrared
behaviour \cite{MMS,ABK}
originated in multi one-loop insertions.

Considering a tadpole Fig.\ref{F1}:e in our case of complex scalar
field we have
\be
\Gamma(p)=\frac{1}{(2\pi)^d}\int d^dk\frac{A+B\cos^2(k\wedge p)}{k^2+m^2}=
\frac{A+\frac{B}{2}}{(2\pi)^d}\int d^dk\frac{1}{k^2+m^2}+\frac{B}{2(2\pi)^d}
\int d^dk\frac{e^{i2k\wedge p}}{k^2+m^2}.
\ee
Integrating this expression over momentum $k$ we obtain
\be
\Gamma(p)=\frac{m^{d-2}}{(4\pi)^{d/2}}(A+\frac{B}{2})\Gamma(1-d/2)+
\frac{B}{(4\pi)^{d/2}}\left[\frac{m}{\xi|\theta p|}\right]^{d/2-1}
K_{d/2-1}(2m\xi|\theta p|).
\ee

If $d=4$ the second term is singular when $p\to 0$.
But in the case $B=0$ this term disappears and hence there is no IR
problem at all.

In conclusion,
we have considered two-coupling  noncommutative
analogue of
$U(1)$-invariant quartic interaction $(\phi^*\phi)^2$ of the complex
scalar field
and shown that renormalizability takes place only in two special
cases, in one of this cases the theory  is free of infrared
divergencies.

\section*{Acknowledgments}

We would like to thank B. Dragovich, P. B. Medvedev, O. A. Rytchkov  and
I. V. Volovich for useful discussions. This work was supported in part
by RFFI grant 99-01-00166  and by grant for the leading scientific
schools.

\end{document}